\documentclass[preprint2]{emulateapj}

\shorttitle{Prompt emission and X-ray plateau}
\shortauthors{Staff et al.}

\begin{document}

   \title{Gamma Ray Burst engine activity within the quark nova scenario:\\
          Prompt emission, X-ray Plateau, and sharp drop-off}

\author{Jan Staff}

\affil{Department of Physics, Purdue University,
525 Northwestern Avenue,
West Lafayette, IN 47907-2036}

\email{jstaff@purdue.edu}

\author{Brian Niebergal and Rachid Ouyed} 

\affil{Department of Physics and Astronomy, University of Calgary,
2500 University Drive NW, Calgary, AB T2N 1N4, Canada}

\begin{abstract} We present a three-stage model for a long GRB inner engine
to explain the prompt gamma ray emission, and interpret recent Swift
satellite observations of early X-ray afterglow plateaus followed by a sharp
drop off or a shallow power law decay. The three stages involves a neutron
star phase, a quark star (QS) and a black hole phase as described in
\citet{staff07}. We find that the QS stage allows for more energy to be
extracted from neutron star to QS conversion as well as from ensuing
accretion onto the QS. The QS accretion phase naturally extends the engine
activity and can account for both the prompt emission and irregular early
X-ray afterglow activity.  Following the accretion phase, the QS can
spin-down by emission of a baryon-free outflow. The magnetar-like magnetic
field strengths resulting from the NS to QS transition provide enough
spin-down energy, for the correct amount of time, to account for the plateau
in the X-ray afterglow. In our model, a sharp drop-off following the plateau
occurs when the QS collapses to a BH during the spin-down, thus shutting-off
the secondary outflow.  We applied our model to GRB 070110 and GRB 060607A
and found that we can consistently account for the energetics and duration
during the prompt and plateau phases.  
\end{abstract}


\section{Introduction}

Observations of Gamma ray bursts (GRBs) by the SWIFT 
satellite \citep{gehrels04} have revealed that many GRBs show a flat segment in their early
X-ray afterglow. This flat segment is often observed to start after about
$10^3$ seconds, and lasts up to $10^5$ seconds. Following the plateau,  some afterglows decay following a
power law with a modest power of about $-1$ to $-2$. However, in some cases
a very sharp drop-off succeeds the plateau.  

In the literature, there are mainly two different explanations for the flattening 
that have been proposed: (i) the 
 refreshed shocks explanation \citep{reesmeszaros98}, where slower shells ejected
during the prompt engine phase catch up with the external shock and
refresh it. The plateau is then followed by a shallow 
decay of power index $-1$ to $-2$ from the
 cooling of the external shock once the shells stop hitting it;
 (ii) Extended engine activity in the form of a secondary outflow 
 \citep[see for example][]{panaitescu07}. 
The secondary outflow explanation
requires that the engine is active for longer than previously expected,
and so if the engine in fact turns off at a later time it can provide an 
explanation for the sharp drop-off in the observed light curve.

In this paper we appeal to a secondary outflow, and propose that it is emitted 
by a quark star (QS; specifically a color-flavor locked strange QS with no crust;
  Ouyed et al. 2005). 
This secondary outflow leads to the observed flattening, and the sharp drop-off
we argue is a consequence of the cessation of the secondary outflow from
the transition of the QS (the GRB inner engine in our model) to a black hole
(BH).  
In a previous paper 
\citep[][hereafter SOB07]{staff07}, a three stage model for long GRBs was
suggested, involving a neutron star (NS)
phase, followed by an accreting QS phase and a
plausible third stage that occurs when the QS accretes enough material
to become a BH. 
The advantages of our model is that, by including a QS phase, we can
account for both energy and duration of the prompt emission, the flattening, 
the sharp drop-off or shallow decay, and the X-ray flaring (see SOB07). 

The secondary outflow is a pair wind due to spin-down from
magnetic braking of the QS (Niebergal et al. 2006). 
Here we suggest that the rotational energy of a rapidly rotating
QS can be used to explain the flattening.  This paper
is organized as follows: In section
\ref{threestagessummary} we briefly describe the
framework of our model. 
In section \ref{sec:spindown} we discuss how the rotational
energy released as the QS spins down due to magnetic braking can
give rise to  flattening in the X-ray afterglow. 
Also, the sharp drop-off is discussed as a signature of the QS turning into a BH.
In section \ref{examples} we apply our model
to GRB 070110 and GRB 060607A. We 
summarize and conclude in section \ref{summary}.

\section{The Three stages}
\label{threestagessummary}

The three stages of the GRB engine described in SOB07 are as follows.
Stage 1 is a (proto-) NS phase, the NS being born in the
collapse of the iron core in an initially massive star. This NS
can collapse to a QS, either by spin-down \citep{staff06} or through
accreting material, thereby increasing its central density sufficiently that
it can form strange quark matter. We suggested that this stage could lead to
a delay between the core collapse and the GRB.  The collapse into a QS, in a
quark nova \citep[QN;][]{ouyed02, keranen05}, releases up to $10^{53}$ ergs that might
help power the explosion of the star. This can possibly explain why GRBs
associated supernovae are often very energetic \citep[see][]{ouyed07, leahy07}.
 If a QS is formed directly in the core collapse,
stage 1 will be bypassed and the process starts from stage 2.

Stage 2 is accretion onto the QS from the surrounding hyperaccreting debris disk,
which is formed from material left over from the collapse of the progenitor. 
This launches a highly variable
ultra-relativistic jet, in which internal shocks can give rise to the gamma
radiation seen in a GRB \citep{ouyed05}. 
This jet will eventually interact with the
surrounding medium creating an external shock that gives rise to the GRB
afterglow.  The afterglow light curve would follow a powerlaw $F_\nu\sim
t^{-(1-2)}$ \citep{sari98}.  However, slower shells can catch up with the
external shock at later times and refresh it. This can lead to a flatter
segment in the X-ray afterglow \citep[e.g.][]{reesmeszaros98} which is 
 commonly seen in GRB afterglows \citep{obrien06, liang07}. 

Stage 3, which occurs if the QS accreted sufficiently that it collapsed to a
BH, is accretion onto the BH which launches
another ultra-relativistic jet, as described in \citet{devilliers05}.  
Interaction between this jet and the QS jet or internal shocks in the BH jet
itself can give rise to flaring commonly seen in the X-ray afterglow of GRBs. 
The BH jet has the potential to be very powerful, so if it
catches up with the external shock a bump might be seen in the light curve.
 The relevant features and emission have been discussed in details in SOB07.
Alternatively, if the QS did not collapse to a BH,
continued accretion onto the QS after the prompt phase might also be able to 
explain X-ray flaring. 

\section{Prompt emission, X-ray Plateau, and sharp drop-off}\label{sec:spindown}

In our model the prompt emission is produced by internal shocks 
in a QS jet launched by hyperaccretion onto a QS \citep{ouyed05}. In this 
section we
will first explain that for the accreting material to be channeled to the
polar cap region, this requires a very high magnetic field. If the QS
survives the accretion and is rapidly rotating, this magnetic field can
then spin the QS down. We will show that a similarly strong magnetic
field is what is needed to get the right spin-down time to explain the
observed flattening.

\subsection{Prompt Emission}

The prompt gamma ray emission
 corresponds to 
synchrotron emission by electrons accelerated in internal shocks in the QS jet.
This jet forms an external shock upon interacting with the surrounding
medium, and synchrotron emission from this external shock is responsible for
the afterglow.

In order to explain the energy observed in the prompt gamma radiation, SOB07
 found that the accretion rate onto the QS must be of the order
$\dot{M}\sim10^{-5}-10^{-3}M_\odot/{\rm s}$. In order to create a jet, the
accretion has to be channeled onto the polar cap. This can occur if the
magnetic radius is at least twice the radius of the star. With the before
mentioned accretion rate, a magnetic field of the order
$B\sim10^{14}-10^{15}G$ \citep[see][]{ouyed05b} is required.\footnote{Recent 
work shows that $10^{15}$ G 
magnetic fields can readily be obtained during QS formation due to the response of quarks to 
the spontaneous magnetization of the gluons \citep[e.g.][and 
references therein]{iwazaki05}. }
It should be noted that this QS jet is much different
than the typical MHD disk wind jets.  A QS jet is created
as the accreting material reaches the surface of the QS, it is converted  into
CFL quark matter, resulting in the creation of a hot spot due to the release
of excess binding energy. This region cools by emitting photons, which
collide with subsequent accreting material, resulting in the ejection of
material with high Lorentz factors \citep[for details, see][]{ouyed05}.

Given that the prompt emission requires such high magnetic fields (because
of the high accretion rates), one has to reconcile this with the plateaus
observed in some light curves at later times. 

A very high magnetic field and a high accretion rate can make the QS
find itself in the propeller regime if it is also spinning very fast
($P\lesssim2$ ms). If the QS is born in the propeller regime,
then we suggest that there will be a delay between the
formation of the QS and the launching of the jet, while the propeller spins
the QS down.

\subsection{Flattening}\label{sec:flattening}

 \citet{panaitescu07} suggested that an outflow, ejected by
the engine after the initial blast, can scatter the forward-shock
synchrotron emission and thereby produce flux that will outshine the primary
one, especially if the outflow is nearly baryon free and highly
relativistic\footnote{An alternative model for
generating the radiation is magnetic reconnection or dissipation processes
in a highly magnetized outflow which was proposed by
\citet[][for the prompt emission]{usov94} and \citet[][for the
afterglow]{gao06}.}. This reflected flux can produce certain light-curve features
such as flares, plateaus, and chromatic breaks.  For this to occur, the
duration of this scattering outflow has to last as long as these observed
features (modulo cosmological time-dilation).

We next show that by using the rotational energy lost from a QS
spinning down, assuming a magnetic field of $10^{15}~{\rm
G}$, a spin-period of
$\sim 2~{\rm ms}$, a characteristic decay time of the order $10^3$ - $10^4$
 seconds  is obtained. The observed flattening in the light-curves of
 certain GRBs can last for several times $10^4$ s 
 and fits well with the duration from the
QS spin-down.

Following the birth of a CFL QS, due the to onset of color superconductivity 
the magnetic flux inside the star 
is forced into a vortex lattice that is aligned with the rotation axis.  
This subsequently forces the magnetic field outside the star to re-structure itself
into a dipole configuration that is aligned with the rotation axis \citep{ouyed_niebergal06}.
Such an aligned
rotator will  spin down by magnetospheric currents escaping
through the light cylinder.  Pair production from magnetic reconnection
supplies these currents \citep{niebergal06} with a corresponding 
 luminosity  given by \citep{shapiroteukolsky83}:
\begin{equation}\label{sdlum}
L=-\dot{E}_{\rm rot}\sim\frac{B^2\Omega^{\left(n+1\right)}R^6}{c^3} \ ,
\end{equation}
where $B$ is the magnetic field at the pole, $R$ is the radius of the star, 
$\Omega$ is the angular rotational frequency of the star, $c$ is the speed of 
light, $n$ is the magnetic braking index.

For an aligned rotator without field decay, the braking index is roughly $n\sim 3$,
however due to magnetic flux expulsion from a CFL QS, the magnetic field decays
as prescribed by \cite{niebergal06}.  This results in an evolution of the luminosity
due to spin-down, which is expressed by the relation,
\begin{eqnarray}\label{eq:qssd_lum}
L &\sim & 3.75\times 10^{48}~{\rm erg~s}^{-1} \left(\frac{B_0}{10^{15}~{\rm G}}\right)^2 
    \left(\frac{2~{\rm ms}}{P_0}\right)^4 \left(1 + \frac{t}{\tau} \right)^{-5/3} \ ,
\end{eqnarray}
where the characteristic spin-down time (in seconds) is,
\begin{equation}\label{eq:charac_time}
\tau = 3.5\times 10^3~{\rm s} \left(\frac{10^{15} {\rm G}}{B_0}\right)^2 
                    \left(\frac{P_0}{2 {\rm ms}}\right)^2 
                    \left(\frac{M_{\rm QS}}{1.4M_{\odot}}\right) \left(\frac{10 {\rm km}}{R_{\rm QS}}\right)^4 \ .
\label{sdtime}
\end{equation}
In the above equations, $M_{\rm QS}$ is the QS mass, $P_{0}$ is the initial spin period, 
and $B_{0}$ is the initial magnetic field strength.  

From Eq.~\ref{eq:qssd_lum} one can see that the luminosity, due to rotational energy
extracted from spin-down of a QS, has a natural break at time $\tau$. 
Thus, if there was a one to one
relationship between spin-down luminosity and observed emission, 
then the power law decay of the observed light-curve 
should change from zero to $-5/3$ after roughly ten thousand seconds.
However, the observed emission might be modified by the forward shock
as discussed in \citet{panaitescu07}.

The energy released from the spin-down of the QS 
is likely to be in the form of an $e^+e^-$ wind. 
Thus, it should be mostly baryon free, since the QS becomes bare immediately
following its birth as it enters the CFL phase \citep[see][]{niebergal06}. 
As in the case of a pulsar, spin-down energy extracted from a QS is mainly
in the equatorial plane. \citet{bucciantini07} performed numerical simulations where they
showed that it is still possible to collimate such equatorial flows into a jet.

A relativistic outflow from the spin-down of a highly-magnetized
\textit{neutron} star has been suggested before as a mechanism to produce
plateaus \citep[for instance in][]{troja07}, however they did not propose a
unified model explaining both the prompt emission and the afterglow
features.  We have here proposed a model that can explain both the prompt
GRB emission and the observed X-ray afterglow features.

\subsection{Sharp vs. Gradual decay}

Eq.~\ref{eq:qssd_lum} naturally gives a break in the engine luminosity at 
$t=\tau$. The engine will also remain active after this break, but the
engine luminosity will gradually decay (with a power law $\sim -5/3$; which
is not necessarily the power law decay in the observed emission). 
In some instances however, it is possible that the QS reaches an unstable
configuration, such that the QS stage is only temporary before the collapse
to a BH.

If the QS collapses to a BH during spin-down, the engine
will likely be shut off. Although the BH is likely to be rapidly
rotating, a disk is necessary in order to extract the rotational energy of a
BH through the Blandford-Znajek mechanism \citep[BZ;][]{bz77}. Only if a
disk has remained around the QS during spin-down or if it is formed
after the formation of the BH, can the BZ mechanism play a role. If
this does not occur, the observed light curve will be generated by the external shock
only after this stage. A sharp drop off will be seen as the light curve
drops from the level given by the spin-down outflow to the level given by the
external shock.

We suggest that in GRB light curves exhibiting plateaus, those possessing a
gradual decay following the plateau are either due to refreshed shocks as
discussed in SOB07 or from spin-down of QSs that have not collapsed to BHs.
If the secondary outflow is responsible for the X-ray afterglow, then 
the external shock can produce the optical afterglow. This scenario might
explain why the optical and X-ray afterglows behave different in some GRBs.

\begin{table}
\begin{center}
\caption{Observed quantities in GRB 070110 and GRB 060607A.
\vspace{-0.3cm}
\label{obstable}}
\begin{tabular}{lllll}
\hline \noalign{\smallskip}
& GRB 070110 & ref & GRB 060607A & ref\\
\hline \noalign{\smallskip} redshift ($z$) & 2.352 & $\dag$ & 3.082 & $\dag$ \\
$E_{iso,X}$ & $1.85\times10^{52}$ ergs & $\dag$ & $6.16\times10^{52}$ ergs & $\dag$\\
$T_{\rm break}$ (engine frame) & $6000$ s & $\ddag$ & $2750$ s & $\ddag$\\
$L_{\rm Obs., iso}$ (during plateau) & $10^{48}$ erg/s  & $\clubsuit$ & $6\times10^{48}$ erg/s & $\spadesuit$ \\
$L_{\rm Eng., 10}$ (during plateau) & $1.5\times10^{46}$ erg/s & $\diamondsuit$ & $1\times10^{47}$ erg/s & $\diamondsuit$ \\
$T_{90}/(1+z)$ & 25.4 s & $\ddag$ & 24.5 s & $\ddag$\\
$E_{\rm \gamma, iso}$ & $2\times10^{52}$ ergs & $\dag$ & $5.2\times10^{52}$ ergs & $\dag$\\
$E_{\rm \gamma,10}$ & $1.0\times10^{50}$ ergs & & $2.5\times10^{50}$ ergs\\
\noalign{\smallskip} \hline
\end{tabular}
\end{center}
References: \\
$\dag$: \citet{liang07} \\
$\ddag$: Calculated using redshift and duration from \citet{liang07} \\
$\clubsuit$: \citet{troja07} \\
$\spadesuit$: Calculated using $E_{\rm iso,X}$, z, and $T_{\rm break}$ from
\citet{liang07}\\
$\diamondsuit$: Observed luminosity corrected for redshift, assuming 10
degrees opening angle\\
\end{table}

\section{Case study}
\label{examples}

In this section we will apply our model to two GRBs, GRB 070110 and GRB
060607A, that both show a flattening followed by a sharp drop off which is
difficult to explain with the external shock.
Some observed properties of both GRBs are summarized in Table~\ref{obstable}.

Based on observations of the duration of the X-ray flattening, we use
Eq.~\ref{eq:charac_time} to estimate the corresponding magnetic field 
strength. We then use Eq.~\ref{eq:qssd_lum} to
find the spin-down luminosity. 
 Both the magnetic field and the spin-down luminosity found this
way are listed in Table~\ref{calculatedtable} which 
 is then compared to observed values (Table~\ref{obstable}). 
 Furthermore, now that we have
an estimate for the magnetic field of the QS, this gives us an estimate
 for  the accretion rate that can be channeled to the polar cap.
 We assume a jet opening angle of about 10 degrees.
 The observed prompt GRB
  emission is then calculated by
  assuming that a combination of accretion efficiency and radiative
 efficiency leads to   $\sim 1\%$ of the 
  total gravitational energy of the accreted material  is converted
  to prompt radiation. As  shown below, for both GRB 070110 and GRB
060607A we find that the magnetic field found based on the duration of the
X-ray flattening consistently and simultaneously  explains the energy of both the GRB itself
and the X-ray flattening.

In our model we know the time at which the QS collapses to a BH (the
time of the steep decay). The calculations above assumed that
this occurred at $t_{\rm collapse}=\tau$. However, it could also
occur at $t_{\rm collapse}<\tau$, which implies that
the magnetic field is weaker than found above. Hence, the magnetic field
found above is the maximum possible magnetic field, and therefore the
spin-down luminosity, accretion rate and prompt gamma ray energy are also
maximum.

\subsection{GRB 060607A and 070110}

\begin{table}[t!]
\begin{center}
\caption{Derived quantities for GRB 070110 and GRB 060607A.
\vspace{-0.3cm}
\label{calculatedtable}}
\begin{tabular}{llll}
\hline \noalign{\smallskip} & GRB 070110 & GRB 060607A \\
\hline \noalign{\smallskip} Maximum magnetic field & $6.8\times10^{14}$ G & $1.0\times10^{15}$ G \\
Spin-down luminosity & $1.6\times10^{48}$ erg/s & $3.6\times10^{48}$ erg/s\\
$\dot{m}_{\rm acc.,max.}$  & $7.5\times10^{-4} M_\odot/{\rm s}$ & $1.6\times10^{-3} M_\odot/{\rm s}$ \\
  $E_{\gamma, 10, max}$ & $3.8\times10^{50}$ ergs & $7.8\times10^{50}$ ergs \\
  $E_{\rm \gamma,iso, max}$ & $7.9\times10^{52}$ ergs & $1.6\times10^{53}$ ergs  \\
\noalign{\smallskip} \hline
\end{tabular}
\end{center}
The maximum magnetic field is calculated using Eq.~\ref{sdtime} assuming
that the QS collapsed to a BH at $t=\tau$ and an initial spin period of 2
ms. The other quantities in this table is calculated based on this maximum
magnetic field.
\end{table}

The QS magnetic field needed to explain the flattening observed 
in GRB 070110 is
$B=6.8\times10^{14}$ G (see Table~\ref{calculatedtable}).
 The corresponding spin-down luminosity is found to be $1.6\times10^{48}$ erg/s.
We can compare this to the observed engine luminosity assuming an opening
angle of 10 degrees for this outflow. If we assume an efficiency of
$10\%$ in converting kinetic energy to photons we see that we have an order
of magnitude more energy than needed. Comparing the observed prompt gamma
ray energy to what we find from the jet launched by the QS, we
again find that the jet energy is higher (by a factor 4) than the observed 
gamma ray energy.

The QS magnetic field needed to explain the flattening 
observed in GRB 060607A is B=$1\times10^{15}$ G (see Table~\ref{calculatedtable}).
 The corresponding spin-down luminosity is found to be $3.6\times10^{48}$
erg/s.
Assuming $10\%$ efficiency in producing X-ray photons, we find 
(as for GRB 070110) that the estimated luminosity is higher than the observed. 
The gamma ray 
energy released during the prompt phase is also higher than the observed
gamma ray energy. 

The higher luminosities can be because the estimate for the
magnetic field is too high, meaning that $\tau$ is larger and that the QS
collapsed to a BH before $t=\tau$. A lower magnetic field implies that
the accretion rate is lower. Alternatively, we have overestimated
the efficiencies, or the opening angle of the outflow is larger.

In GRB06067A there are several X-ray flares observed until about 300 seconds
  (about 75 seconds when corrected for redshift). If we explain
these flares by accretion onto the QS as well, that means that the accretion
process lasts for about $75$ seconds. The derived accretion rates
imply the necessity of a debris disk with a mass of the order of  $\sim 10^{-1}M_\odot$,
which is reasonable since the QN goes off inside a collapsar,
 where such a large fall-back disk is in principle allowed.

\section{Summary \& Conclusion}
\label{summary}

We have presented a model to explain the flattening and occasional
sharp drop-off seen in X-ray afterglows of some GRBs. Our model 
 borrows the framework of the 3 stage model presented in SOB07 which
 makes use of an intermediate QS stage between the NS and the BH.
  By appealing to a secondary outflow, from the QS
 spin-down due to magnetic braking, our model seems to explain 
  the GRB itself (i.e. prompt emission), the observed flat segment (i.e.
   plateau), and the subsequent sharp or gradual decay following the
   plateau.  The sharp or gradual decay depends on whether the QS
    collapses to a BH or not during spin-down.
    During spin-down, a break will be seen after a characteristic time
$\tau$ given by Eq.~\ref{eq:charac_time} followed by a power law with power
of $-5/3$ to $-3$ \citep{panaitescu07}. A very sharp drop-off will be seen
if the QS collapses to a BH during spin-down. 

We note that, if there was a way for launching ultrarelativistic jets from
accretion onto NSs, then it would be tempting to not include the
QS phase in our model and appeal only to NS to BH transition.
However, we are not aware of any such mechanism for launching an 
ultrarelativistic jet from accretion onto a NS, and from an energetics perspective
it seems unlikely.  Hence, the additional energy available from converting 
hadronic to strange quark matter and during accretion onto the QS seems crucial
in explaining the nature of GRBs.

In addition to an energetics point of view, the most important
benefits of our GRB model involving a QS stage are:
(i) the QS offers an additional stage that allows for more energy
  to be extracted from the conversion from NS to QS as well
  as from accretion. Also, additional energy is released
   as the QS quickly evolves from a non-aligned
   to an aligned rotator following its birth with up
   to $10^{47}$ ergs released in a few seconds \citep{ouyed_niebergal06}. As such, the QS phase extends
   the engine activity and so can account for both
  the prompt emission and  irregular X-ray afterglow activity;
 (ii) a natural amplification of the NS magnetic field to
  $10^{14}$-$10^{15}$ G during the transition to the QS \citep{iwazaki05}.
   Such high strengths gives the correct spin down time
  to for  the plateau;  
 (iii) since QS in the CFL phase
 might not  have a crust, the spin down energy 
 will most likely be extracted as an $e^+e^-$ fireball with very little
baryon contamination \citep[see discussion in][]{niebergal06}.
\citet{panaitescu07} favors a baryon free  secondary outflow to explain
  the plateau.

\acknowledgments{
 We thank Y. Fan and D. Xu for comments.
}

\end{document}